\documentclass[aps,prd,showpacs,nofootinbib,floats,floatfix,
preprintnumbers,onecolumn]{revtex4}
\usepackage{bm}
\usepackage{latexsym}
\usepackage{dcolumn}
\usepackage{amsfonts,amssymb}
\usepackage{graphicx,epsfig}
\def\beq{\begin{equation}}
\def\eeq{\end{equation}}
\def\bea{\begin{eqnarray}}
\def\eea{\end{eqnarray}}
\def\benu{\begin{enumerate}}
\def\eenu{\end{enumerate}}
\def\nn{\nonumber}

\def\l{\left}
\def\r{\right}
\def\lp{L_{_{\rm P}}}
\reversemarginpar

\begin{document}
%\preprint{arXiv:0712xxxx [gr-qc]}
\title{Quantum gravitational corrections to the stress-energy\\
tensor around the rotating BTZ black hole}
\author{Dawood A.~Kothawala}\email[]{E-mail: dawood@iucaa.ernet.in}
\affiliation{IUCAA, Post Bag 4, Ganeshkhind, Pune 411 007, India.}
\author{S.~Shankaranarayanan}\email[]{E-mail: Shanki.Subramaniam@port.ac.uk}
\affiliation{Institute of Cosmology and Gravitation, University of
Portsmouth, Mercantile House, Portsmouth~P01 2EG, U.K..}
\author{L.~Sriramkumar}\email[]{E-mail: sriram@mri.ernet.in}
\affiliation{Harish-Chandra Research Institute, Chhatnag Road,
Jhunsi, Allahabad~211~019, India.}
\date{\today}
%%%%%%%%%%%%%%%%%%%%%%%%%%%%%%%%%%%%%%%%%%%%%%%%%%%%%%%%%%%%%%%%%%%%%%%%%%%%%%%
\begin{abstract}
Modes emerging out of a collapsing black hole are red-shifted to
such an extent that Hawking radiation at future null infinity
consists of modes that have energies beyond the Planck scale at
past null infinity. This indicates that physics at the Planck
scale may modify the spectrum of Hawking radiation and the
associated stress-energy tensor of the quantum field. Recently, it
has been shown that, the T-duality symmetry of string fluctuations
along compact extra dimensions leads to a modification of the
standard propagator of point particles in quantum field theory. At
low energies (when compared to the string scale), the modified
propagator is found to behave as though the spacetime possesses a
minimal length, say, $\lp$, which we shall assume to be of the
order of the Planck length. We utilize the duality approach to
evaluate the modified propagator around the rotating
Banados-Teitelboim-Zanelli black hole and show that the propagator
is finite in the coincident limit. We compute the stress-energy
tensor associated with the modified Green's function and
illustrate graphically that the quantum gravitational corrections
turn out to be negligibly small. We conclude by briefly commenting
on the results we have obtained.
\end{abstract}
%04.70.Dy Quantum aspects of black holes, evaporation, thermodynamics
%04.62.+v Quantum fields in curved spacetime
%04.60.Kz Lower dimensional models; minisuperspace models
%04.60.-m Quantum gravity
\maketitle
\date{\today}

\reversemarginpar

%%%%%%%%%%%%%%%%%%%%%%%%%%%%%%%%%%%%%%%%%%%%%%%%%%%%%%%%%%%%%%%%%%%%%%%%%%%%%%%
\section{Why do we need to consider Planck scale physics?}

Hawking radiation primarily arises due to the asymmetry in the
extent of the redshift and the blueshift of the modes of a quantum
field as they propagate through matter that is collapsing
gravitationally and, eventually, goes on to become a black
hole~\cite{hawking-op}. Consider a typical mode that constitutes
Hawking radiation at the future null infinity (i.e. at ${\cal
I}^{+}$), say, the mode where the intensity of the radiation is
the maximum and whose wavelength can be identified, for instance,
using Wein's law. When one traces such a mode back to the past
null infinity (i.e. to ${\cal I}^{-}$) where the initial
conditions are imposed on the quantum field, one finds that the
energy of the mode turns out to be way beyond the Planck scale.
(This feature seems to have been originally noticed in
Ref.~\cite{wald-1976}; in this context, also see
Ref.~\cite{wald-1984}.) As Hawking radiation mostly consists of
modes that leave the future event horizon just before its
formation, such a phenomenon essentially occurs due to the
enormous red-shifting of the modes near the horizon. This behavior
then raises the question as to whether the Planck scale effects
will modify the spectrum of Hawking radiation and the associated
stress-energy tensor of the quantum field.

There has been a sufficient amount of effort in the literature
towards understanding the effects of Planck scale physics on
Hawking radiation (for the earliest discussions, see
Refs.~\cite{jacobson-1993-99,brout-1995-99,
hambli-1996,parentani-1999-2001}, and, for relatively recent
efforts, see Refs.~\cite{casadio-2006,agullo-2007}). In the
absence of a workable quantum theory of gravity, to study the
Planck scale effects, most of these efforts (apart from one
notable exception, see Ref.~\cite{hambli-1996}) consider
phenomenological models constructed by hand---models which
purportedly contain one or more features of the actual effective
theory obtained by integrating out the gravitational degrees of
freedom. These models either introduce new features in the
standard dispersion
relation~\cite{jacobson-1993-99,brout-1995-99}, or work with a
classical fluctuating geometry~\cite{parentani-1999-2001}, or
assume that the spacetime coordinates are
non-commutative~\cite{casadio-2006}. Often---though, we should
hasten to add---but, not always, these high energy models do not
preserve local Lorentz invariance. Moreover, some of them either
consider a simpler model of Hawking radiation (say, the popular
model of a moving mirror in flat space-time~\cite{casadio-2006})
rather than the actual situation, or just consider the spherically
symmetric (i.e. the $\ell=0$) mode in higher dimensional cases,
which, essentially, reduces to studying the effects in
$(1+1)$-dimensions. Our aim in this work, is to use a locally
Lorentz invariant approach to evaluate the Planck scale
modifications to stress-energy tensor around a black hole without
resorting to such approximations.

General arguments based on the merging of essential concepts from
general relativity and quantum mechanics seem to indicate that it
may not be possible to probe spacetime intervals smaller than the
Planck length, say, $L_{_{\rm P}}$ (see, for example,
Ref.~\cite{paddy-1987}). An approach that introduces such a
`zero-point length' into standard quantum field theory while
preserving local Lorentz invariance is the so-called principle of
path integral duality (for the original discussion, see
Refs.~\cite{paddy-1997-98}; for various applications of the
approach, see Refs.~\cite{srini-1998,shanki-2001,sriram-2006}).
Interestingly, it has been shown that, at low energies (when
compared to the string scale), the modified propagator of matter
fields obtained through such an approach is equivalent to taking
into account the string fluctuations propagating along compact
extra
dimensions~\cite{smailagic-2003,spallucci-2005,fontanini-2006}.
Effectively, the path integral duality approach can be said to
provide a prescription to evaluate the modified Green's function
of a free quantum field that is propagating in a given classical
background (for further details, see the following section). In
this work, we shall use this prescription to evaluate the modified
Green's function and the Planck scale corrections to the
stress-energy tensor around a specific black hole.

In $(3+1)$-dimensions, it proves to be difficult to evaluate the
two-point function exactly around even the simplest of black
holes. As a result, the stress-energy tensor of a quantum field
around, say, the Schwarzschild black hole has been evaluated only
under an approximation (see, for example,
Refs.~\cite{page-1982,campos-1998,phillips-2003}). In contrast,
the spacetime around the $(2+1)$-dimensional, rotating
Banados-Teitelboim-Zanelli (BTZ, hereafter) black
hole~\cite{banados-1992-93,carlip-1995,carlip-1998} provides a
situation wherein it is possible to calculate the two-point
function in a closed
form~\cite{steif-1994,lifschytz-1994,shiraishi-1994,ichinose-1995,
mann-1997,bystenko-1998,binosi-1999}. We shall utilize this
feature to compute the duality modified propagator and the
corresponding Planck scale corrections to the standard
stress-energy tensor around the rotating BTZ black hole.

The remainder of this paper is organized as follows. In the
following section, we shall briefly outline as to how the
principle of path integral duality modifies the two-point function
of a free quantum field evolving in a given spacetime. In
Section~\ref{sec:mGfn}, we evaluate the modified propagator around
the rotating BTZ black hole, and show that the path integral
duality approach regulates the ultra-violet behavior of the
two-point function. In Section~\ref{sec:mst}, we evaluate the
stress-energy tensor associated with the modified two-point
function and graphically illustrate the form of the Planck scale
modifications to the stress-energy tensor. Finally, in
Section~\ref{sec:summary}, we close with a brief discussion on the
results we have obtained.

Before we proceed, a few words on our notations and conventions
are in order. We shall work in $(2+1)$-dimensions and adopt the
metric signature of $(-, +, +)$. Also, for convenience, we shall
denote the set of three coordinates $x^{\mu}$ as ${\tilde x}$, and
use natural units such that $G=\hbar =c=1$.

%%%%%%%%%%%%%%%%%%%%%%%%%%%%%%%%%%%%%%%%%%%%%%%%%%%%%%%%%%%%%%%%%%%%%%%%%%%%%%%

\section{String fluctuations, duality and the modified
Green's function}

Recently, it was shown that, when the fluctuations of closed
strings along compact extra dimensions are taken into account, one
arrives at an effective propagator for point particles that is
regular at high energies~\cite{fontanini-2006}. It was found that,
the $T$-duality symmetry between the topologically non-trivial
excitations and the winding modes of the strings around the
compact dimensions leads to a modified propagator in the Minkowski
vacuum wherein the original spacetime interval $({\tilde x} -
{\tilde x}')^2$ between the two spacetime events ${\tilde x}$ and
${\tilde x}'$ is replaced by $[({\tilde x} - {\tilde x}')^2 +
\lp^2]$, where, as we mentioned above, $\lp$ denotes the Planck
length. Thus, effectively, the string fluctuations introduce a
zero point length $\lp$ into the standard field theory.

A similar modification of the Minkowski propagator has been
obtained earlier by invoking the principle of path integral
duality~\cite{paddy-1997-98}. Recall that, in standard quantum
field theory, the path integral amplitude for a path connecting
events~${\tilde x}$ and~${\tilde x'}$ in a given spacetime is
proportional to the proper length, say, ${\cal R}({\tilde
x},{\tilde x'})$, between the two events. The duality principle
proposes that the path integral amplitude should be invariant
under the transformation ${\cal R} \rightarrow \l(L_{_{\rm
P}}^2/{\cal R}\r)$. Operationally, it turns out to be convenient
to express the effect of the duality approach and the string
fluctuations on the propagator in terms of the Schwinger's proper
time formulation~\cite{paddy-1997-98}. One finds that, in such a
formulation, the string fluctuations and the duality approach
effectively modify the weightage given to a point particle of mass
$m$, from $\exp\; -(i\, m^2\,s)$ to $\exp\; -[i\, (m^2\, s -
(\lp^2/4\,s))]$.

Consider a free scalar field of mass~$m$ that is propagating in a
given classical gravitational background described by the metric
tensor~$g_{\mu\nu}$. Let us further assume that the field is
non-minimally coupled to gravity. In Schwinger's proper time
formalism, the two-point function corresponding to such a scalar
field can be expressed as~\cite{schwinger-1951,dewitt-1975} \beq
G({\tilde x},{\tilde x'}) =i\,\int\limits_0^{\infty}ds\; K({\tilde
x}, {\tilde x'}; s), \label{eq:Gfn} \eeq where $K({\tilde x},
{\tilde x'}; s)$ is defined as \beq K({\tilde x}, {\tilde x'}; s)
\equiv \langle {\tilde x}\vert\, e^{i\, \l({\widehat \Box}-m^2
-\xi\, R\r)\, s}\, \vert {\tilde x'}\rangle, \label{eq:k} \eeq and
$R$ is the curvature of the background spacetime with $\xi$ being
the coefficient of the non-minimal coupling. In other words, the
quantity $K({\tilde x}, {\tilde x'}; s)$ is the path integral
amplitude for a quantum mechanical system described by the
following Hamiltonian: \beq {\widehat H} = -\l({\widehat
\Box}-m^{2}-\xi\, R\r) \equiv -\frac{1}{\sqrt{-g}}\,
\partial_{\mu}\l(\sqrt{-g}\, g^{\mu\nu}\, \partial_{\nu}\r)
+m^2+ \xi\, R. \label{eq:Hk} \eeq As we mentioned, the effects due
to path integral duality or, equivalently, the string
fluctuations, correspond to modifying the
expression~(\ref{eq:Gfn}) above for the two-point function
to~\cite{paddy-1997-98,fontanini-2006} \beq G^{\rm M}({\tilde
x},{\tilde x'}) =i\,\int\limits_0^{\infty}ds\;\, e^{i L_{_{\rm
P}}^2/4\,s}\;\, K({\tilde x}, {\tilde x'}; s).\label{eq:mGfn} \eeq
In the following section, we shall make use of this prescription
to evaluate the modified two-point function around the rotating
BTZ black hole.

%%%%%%%%%%%%%%%%%%%%%%%%%%%%%%%%%%%%%%%%%%%%%%%%%%%%%%%%%%%%%%%%%%%%%%%%%%%%%%%

\section{The modified Green's function around the rotating
BTZ black hole}\label{sec:mGfn}

The rotating BTZ black hole is an axially symmetric vacuum
solution of the Einstein's equations in three-dimensional anti-de
Sitter spacetime (AdS$_{3}$, hereafter). It can be conveniently
represented as AdS$_3$ identified under a discrete subgroup of its
isometry group~\cite{banados-1992-93,carlip-1995,carlip-1998}.

Recall that AdS$_3$ is a maximally symmetric space sourced by a
{\it negative}\/ cosmological constant, say, $-\ell^{-2}$. AdS$_3$
can be described by the line-element \beq ds^2  = - \l(\frac{{\hat
r}^2}{\ell^2} -1\r)\, d{\hat t}^2 + \l(\frac{{\hat
r}^2}{\ell^2}-1\r)^{-1}\, d{\hat r}^2 + {\hat r}^2\, d{\hat
\phi}^2,\label{eq:AdS} \eeq where  $-\infty < ({\hat t}, {\hat
\phi}) <\infty$ and $0<{\hat r} <\infty$. It is important to note
that the coordinate ${\hat \phi}$ in the above line-element has an
infinite range and, hence, is {\it not}\/ periodic. The rotating
BTZ black hole of mass $M$ and angular momentum~$J$ can be
obtained from the above AdS$_3$ line-element by making the
coordinates ${\hat t}$ and ${\hat \phi}$ suitably periodic as
follows~\cite{banados-1992-93,carlip-1995,carlip-1998}: \beq
({\hat t}, {\hat r}, {\hat \phi}) \equiv \l[({\hat t}-2\, \pi\,
n\, \ell\, \alpha_{_{-}}), {\hat r}, ({\hat \phi}+ 2\, \pi\,
n\,\alpha_{_{+}})\r],\label{eq:pi} \eeq where, $n$ is an integer,
and the quantities $\alpha_{\pm}$ are given by \beq
\alpha_{_{\pm}} = \l(\frac{1}{2}\r)\, \l[\sqrt{M+(J/\ell)} \pm
\sqrt{M-(J/\ell)} \r]. \label{eq:defAlpha} \eeq On redefining the
coordinates $({\hat t}, {\hat r},{\hat \phi})$ as follows~(see,
for instance, Ref.~\cite{steif-1994}): \beq {\hat t}=
\l(\alpha_{_{+}}\, t - \alpha_{_{-}}\, \ell\, \phi\r), \quad {\hat
\phi}= \l[\alpha_{_{+}}\, \phi - (\alpha_{_{-}}\, t/\ell)\r],
\quad{\rm and}\quad {\hat r^2}=\l(\frac{r^2 - (\alpha_{_{-}}\,
\ell)^2 }{\alpha_{_{+}}^2- \alpha_{_{-}}^2}\r), \label{eq:ct1}
\eeq we obtain the metric around the rotating BTZ black hole to
be~\cite{banados-1992-93,carlip-1995,carlip-1998} \beq ds^2 = -
\l(\frac{r^2}{\ell^2} -M\r)\, dt^2 - J\, dt\, d\phi +
\l(\frac{r^2}{\ell^2} - M + \frac{J^2}{4\, r^2} \r)^{-1}\, dr^2 +
r^2\, d\phi^2\label{eq:BTZ} \eeq with $-\infty < t < \infty$ and
$0< r <\infty$. Using the relations~(\ref{eq:ct1}), it is easy to
show that, under the periodicity conditions~(\ref{eq:pi}), while
$t\to t$, $\phi\to (\phi+2\,\pi\, n)$. Evidently, $\phi$ is a
genuine angular coordinate that can be restricted to the domain
$0< \phi \le 2\pi$. The rotating BTZ black hole
solution~(\ref{eq:BTZ}) contains two horizons, and the locations
of the outer ($r_{_{+}}$) and inner ($r_{_{-}}$) horizons are
given by \beq r_{_{\pm}}=\l(\alpha_{_{\pm}}\, \ell\r) =
\l(\frac{\ell}{2}\r)\, \l[\sqrt{M+(J/\ell)} \pm \sqrt{M-(J/\ell)}
\r]. \eeq Note that, when $J=0$, $r_{_{-}}$ vanishes, and
$r_{_{+}} =(\sqrt{M}\, \ell)$.

It is clear from the above discussion that, if we can evaluate the
Green's function, say, $G_{_{\rm AdS}}({\tilde x},{\tilde x}')$,
in AdS$_{3}$, then the Green's function, say, $G_{_{\rm
BTZ}}({\tilde x},{\tilde x}')$ around the rotating BTZ black hole
can be obtained on imposing by hand the periodicity
condition~(\ref{eq:pi}), and transforming to the black hole
coordinates using the
relations~(\ref{eq:ct1})~\cite{steif-1994,lifschytz-1994,
shiraishi-1994,ichinose-1995,mann-1997,bystenko-1998,binosi-1999}.
Therefore, we need to first evaluate the Green's function in
AdS$_{3}$. We shall do so by evaluating the quantum mechanical
kernel~$K({\tilde x},{\tilde x}';s)$ and using the Schwinger's
proper time expression~(\ref{eq:Gfn}) to arrive at the Green's
function.

In order to calculate the kernel~$K({\tilde x},{\tilde x}';s)$ in
AdS$_{3}$, it turns out to be more convenient to work with the
following set of coordinates~\cite{carlip-1995}: \bea
x_{2}&=&{\hat r}^{-1}\; \sqrt{\l({\hat r}^{2}-\ell^2\r)}\;\,
e^{\phi}\; {\rm sinh}\,({\hat t}/\ell),\nn\\
y&=&\l(\ell/{\hat r}\r)\; e^{\phi},\\
x_{1}&=&{\hat r}^{-1}\; \sqrt{\l({\hat r}^{2}-\ell^2\r)}\;\,
e^{\phi}\; {\rm cosh}\,({\hat t}/\ell).\nn \eea In terms of these
new coordinates, the AdS$_{3}$ line-element~(\ref{eq:AdS}) reduces
to \beq ds^{2}=(\ell^2/y^2)\, \l(-dx_{2}^2+dy^2+dx_{1}^2\r),
\label{eq:cfs} \eeq which, evidently, is conformal to flat
spacetime. The kernel of a massive and non-minimally coupled
scalar field propagating in the above conformally flat
line-element can be easily evaluated using the method of spectral
decomposition~(see, for instance, the Appendix in
Ref.~\cite{binosi-1999}). We find that the kernel can be expressed
as~\cite{mann-1997,bystenko-1998,binosi-1999} \beq K({\tilde
x},{\tilde x}';s) = \l(\frac{1}{(4 \pi\,i\, s)^{3/2}}\r)\;
\l(\frac{\sigma({\tilde x},{\tilde x}')/\ell}{\sinh
\l[\sigma({\tilde x},{\tilde x}')/\ell \r]}\r)\; \exp\, \l[i\,
(\sigma^{2}({\tilde x},{\tilde x}')/4\, s) - i\, (b\, s
/\ell^2)\r],\label{eq:kAdS} \eeq where $b = \l[1 + (m\, \ell)^{2}+
\xi\, R\, \ell^2\r]$, with $R= -(6/\ell^2)$ being the Ricci scalar
(a constant in AdS$_{3}$). The quantity $\sigma({\tilde x},
{\tilde x}')$ denotes the geodesic distance between the two points
${\tilde x}$ and ${\tilde x}'$ in AdS$_{3}$ and is given by \beq
{\rm sinh}\,\l[\sigma({\tilde x},{\tilde x}')/2\,\ell\r]
=\l[\l(-\l(x_{2}-x_{2}'\r)^{2}+\l(y-y'\r)^{2}
+\l(x_{1}-x_{1}'\r)^{2}\r)/4\, y\,y'\r]^{1/2}. \eeq (It may be
useful to note that the geodesic distance in AdS$_{3}$, viz.
$\sigma({\tilde x}, {\tilde x}')$, can be conveniently expressed
in terms of the chordal distance between the two points in the
embedding space.) In terms of the original coordinates $({\hat
t},{\hat r}, {\hat \phi})$, the quantity $\sigma({\tilde
x},{\tilde x}')$ turns out to be \bea &
&\!\!\!\!\!\!\!\!\!\!\!\!\!\!\!\!
{\rm sinh}\l[\sigma({\tilde x},{\tilde x}')/2 \ell\r]\nn\\
& &\;=\l(\sqrt{2}\, \ell\r)^{-1} \biggl[-\, \sqrt{\l({\hat
r}^2-\ell^2\r)\,\l({\hat r}'^2-\ell^2\r)}\;\;
{\rm cosh}\l[\l({\hat t}-{\hat t}'\r)/\ell\r]\nn\\
& &\qquad\qquad\qquad\qquad\qquad\qquad\qquad\qquad
-\;\ell^{2}+{\hat r}\;{\hat r}'\; {\rm cosh}\l({\hat \phi}-{\hat
\phi}'\r)\biggr]^{1/2}.\quad \label{eq:sigma} \eea On using the
expression~(\ref{eq:Gfn}), the Green's function corresponding to
the kernel~(\ref{eq:kAdS}) can then be immediately evaluated to
be~\cite{mann-1997,bystenko-1998,binosi-1999} \beq
G_{_{\rm AdS}}({\tilde x},{\tilde x}')\nn\\
=\l(\frac{1}{4\,\pi\, \ell}\r)\; \l(\frac{1}{\sinh
\l[\sigma({\tilde x},{\tilde x}')/\ell\r]}\r)\;
\exp{-\l[\sqrt{b}\;\sigma({\tilde x},{\tilde x}')/\ell\r]}.
\label{eq:GfnAdS} \eeq We should mention here that this Green's
function corresponds to a particular choice of boundary condition
(actually, the Dirichlet
condition~\cite{mann-1997,bystenko-1998,binosi-1999}) that is
required to be imposed at spatial infinity in AdS spacetimes (see,
for example, Refs.~\cite{ads}).

The Green's function around the rotating BTZ black hole can now be
obtained by imposing the periodicity condition~(\ref{eq:pi}) in
the AdS$_{3}$ Green's function~(\ref{eq:GfnAdS}), and transforming
into the black hole coordinates using the
relations~(\ref{eq:ct1}). The BTZ Green's function can be
expressed
as~\cite{steif-1994,lifschytz-1994,shiraishi-1994,ichinose-1995,
mann-1997,bystenko-1998,binosi-1999} \beq G_{_{\rm BTZ}}({\tilde
x},{\tilde x}') =\l(\frac{1}{4\,\pi\,\ell}\r)\;
\sum\limits_{n=-\infty}^{\infty} \l(\frac{1}{\sinh
\l[\sigma_{n}({\tilde x},{\tilde x}')/\ell \r]}\r)\,
\exp{-\l[\sqrt{b}\;\sigma_{n}({\tilde x},{\tilde x}')/\ell\r]},
\label{eq:GfnBTZ} \eeq with $\sigma_{n}({\tilde x},{\tilde x}')$
given by \bea & &\!\!\!\!\!\!\!\!\!\!\!\!\!\!\!\!\!
{\rm sinh}\l[\sigma_{n}({\tilde x},{\tilde x}')/2\,\ell\r]\nn\\
& &=\l[2\, \ell^2\, \l(\alpha_{_{+}}^{2}
-\alpha_{_{-}}^2\r)\r]^{-1/2}\;\nn\\
& &\qquad\quad
\times\,\biggl(-\,\sqrt{\l[r^2-(\alpha_{_{+}}\,\ell)^2\r]\,
\l[r'^2-(\alpha_{_{+}}\,\ell)^2\r]}\nn\\
& &\qquad\qquad\qquad\quad \times\; {\rm
cosh}\l[(\alpha_{_{+}}/\ell)\; (t-t')
-\alpha_{_{-}}\, (\phi-\phi'+2\,\pi\,n)\r]\nn\\
& &\qquad\qquad\quad
-\, \l(\alpha_{_{+}}^{2}-\alpha_{_{-}}^2\r)\, \ell^{2}\nn\\
& & \qquad\qquad\quad +\,\sqrt{\l[r^2-(\alpha_{_{-}}\,
\ell)^2\r]\,
\l[r'^2-(\alpha_{_{-}}\,\ell)^2\r]}\nn\\
& &\qquad\qquad\qquad\quad \times\; {\rm
cosh}\l[(\alpha_{_{-}}/\ell)\; (t-t')-
\alpha_{_{+}}\,\l(\phi-\phi'+2\pi\, n\r)\r]\biggr)^{1/2}.
\label{eq:sigman} \eea The time translational invariance clearly
indicates that the above Green's function corresponds to the
Hawking-Hartle state of the black hole (see, for instance,
Ref.~\cite{birrell-1982}).

The duality modified Green's function in AdS$_{3}$ can now be
obtained by substituting the kernel~(\ref{eq:kAdS}) in the
expression~(\ref{eq:mGfn}) and carrying out the integral over $s$.
We obtain the duality modified Green's function in AdS$_{3}$ to be
\bea \!\!\!\!\!\!\!\!\!\!\!\!\!\!\!\! G_{_{\rm AdS}}^{\rm
M}({\tilde x},{\tilde x}') &=& \l(\frac{1}{4\, \pi\, \ell}\r)\;
\l(\frac{\sigma({\tilde x},{\tilde x}')}{\sinh \l[\sigma({\tilde
x}, {\tilde x}')/\ell \r]}\r)\; \l(\frac{1}{\sqrt{\sigma^2({\tilde
x},{\tilde x}')
+ \lp^2}}\r)\nn\\
& &\qquad\qquad\qquad\qquad\qquad\quad \times\;
\exp{-\l(\sqrt{(b/\ell^2)\; [\sigma^2({\tilde x},{\tilde x}') +
\lp^2]}\,\r)} \label{eq:mGfnAdS} \eea with $\sigma({\tilde
x},{\tilde x}')$ given by Eq.~(\ref{eq:sigma}). In the coincidence
limit (i.e. when ${\tilde x} \to {\tilde x}'$), $\sigma({\tilde
x},{\tilde x}') \rightarrow 0$, and the above modified Green's
function reduces to \beq G_{_{\rm AdS}}^{\rm M}({\tilde x},{\tilde
x}') =\l(\frac{1}{4\,\pi\,L_{_{\rm P}}}\r)\; \exp{-\l(\sqrt{b}\;
L_{_{\rm P}} / \ell\r)}. \eeq Note that the Green's function is
finite in the coincident limit independent of the nature of the
coupling and the mass of the scalar field. This behavior clearly
illustrates that the string theory inspired modification regulates
the theory at the Planck scale.

The modified Green's function around the rotating BTZ black hole
can be obtained from the above modified Green's function in
AdS$_{3}$ as in the standard case. It can be expressed as \bea
\!\!\!\!\!\!\!\!\!\!\!\!\!\!\!\! G_{_{\rm BTZ}}^{\rm M}({\tilde
x},{\tilde x}') &=& \l(\frac{1}{4\pi\, \ell}\r)\;
\sum\limits_{n=-\infty}^{\infty} \l(\frac{\sigma_{n}({\tilde x},
{\tilde x}')}{\sinh \l[\sigma_{n}({\tilde x},{\tilde
x}')/\ell\r]}\r)\; \l(\frac{1}{\sqrt{\sigma_{n}^2({\tilde
x},{\tilde x}')
+\lp^2}}\r)\nn\\
& &\qquad\qquad\qquad\qquad\qquad\;
\times\;\exp{-\l(\sqrt{(b/\ell^2)\;\, [\sigma_{n}^2({\tilde
x},{\tilde x}') + \lp^2]}\,\r)} \label{eq:mGfnBTZ} \eea with
$\sigma_{n}({\tilde x},{\tilde x}')$ given by
Eq.~(\ref{eq:sigman}). The $n\ne 0$ terms are finite in the
coincident limit even in the standard (i.e. unmodified) case,
while the $n=0$ term corresponds to AdS$_{3}$. Obviously, the
modified Green's function around the rotating BTZ black hole is
finite in the coincident limit as well.

%%%%%%%%%%%%%%%%%%%%%%%%%%%%%%%%%%%%%%%%%%%%%%%%%%%%%%%%%%%%%%%%%%%%%%%%%%%%%%%

\section{Planck scale modifications to the stress-energy
tensor}\label{sec:mst}

In this section, we shall first evaluate the standard
stress-energy tensor around the rotating BTZ black hole using the
two-point function~(\ref{eq:GfnBTZ}). We shall then calculate the
Planck scale modifications to the stress-energy tensor using the
modified Green's function~(\ref{eq:mGfnBTZ}). For convenience in
calculation, we shall restrict ourselves to the case of a massless
and conformally coupled scalar field [i.e. when $m=0$, $\xi=(1/8)$
and, hence, $b=(1/4)$]. In such a situation, given the symmetric
Green's function, say, $G({\tilde x}_{1}, {\tilde x}_{2})$, the
corresponding mean value of the stress-energy tensor can be
expressed as~\cite{steif-1994,lifschytz-1994,shiraishi-1994} \beq
\langle {\hat T}_{\mu\nu}\rangle = \lim_{2\to 1}\; {\cal
T}_{\mu\nu}^{(1,2)}\; G({\tilde x}_{1}, {\tilde
x}_{2}).\label{eq:set} \eeq The quantity ${\cal
T}_{\mu\nu}^{(1,2)}$ is a differential operator and is given by
\bea {\cal T}_{\mu\nu}^{(1,2)} &=&\biggl[\l({3\over 4}\r)\,
\l(\nabla^{1}_{\mu}\, \nabla^{2}_{\nu}\r) -\l({1\over 4}\r)\,
g_{\mu\nu}\;  g^{\alpha\beta}\,
\l(\nabla^1_{\alpha}\,  \nabla^{2}_{\beta}\r)\nn\\
& &\qquad\qquad\qquad\quad -\,\l(\frac{1}{4}\r)\,
\l(\nabla^{1}_{\mu}\,  \nabla^{1}_{\nu}\r) +\l(\frac{R}{96}\r)\,
g_{\mu\nu}\biggr], \eea where the covariant derivatives
$\nabla_{\mu}^1$ and $\nabla_{\mu}^{2}$ act on the points ${\tilde
x}_{1}$ and ${\tilde x}_{2}$, respectively, and $R$ denotes the
scalar curvature of the background spacetime. It should be
mentioned here that the Green's functions~(\ref{eq:GfnBTZ})
and~(\ref{eq:mGfnBTZ}) and, hence, the corresponding stress-energy
tensors are valid in the domain $r> r_{_{+}}$.

%%%%%%%%%%%%%%%%%%%%%%%%%%%%%%%%%%%%%%%%%%%%%%%%%%%%%%%%%%%%%%%%%%%%%%%%%%%%%%%

\subsection{The standard (unmodified) stress-energy tensor}

%%%%%%%%%%%%%%%%%%%%%%%%%%%%%%%%%%%%%%%%%%%%%%%%%%%%%%%%%%%%%%%%%%%%%%%%%%%%%%%

Before we proceed to evaluate the Planck scale modifications, let
us evaluate the stress-energy tensor in the standard case. Also,
let us first consider the simpler situation of the non-rotating
BTZ black hole. In such a case, $J=0$, so that $\alpha_{_{+}}=M$
and $\alpha_{_{-}}=0$. For the case of the Dirichlet boundary
condition imposed at spatial infinity, the mean value of the
stress-energy tensor associated with a massless and conformally
coupled scalar field can be obtained by substituting the Green's
function~(\ref{eq:GfnBTZ}) [with $b=(1/4)$] in the
expression~(\ref{eq:set}) above. We find that the resulting
stress-energy tensor is diagonal and can be expressed
as~\cite{lifschytz-1994,shiraishi-1994} \bea & &\!\! \langle {\hat
T}^{\mu}_{\nu}\rangle
=\l(\frac{M^{3/2}}{32\, \pi\, r^{3}}\r)\nn\\
& &\qquad\qquad\times\; \sum\limits_{n=1}^{\infty}\;\;
\l[2+\l(1-\l[1+\l(\frac{M\, \ell^{2}}{r^{2}\, {\rm sinh}^{2}\,
(\sqrt{M}\, \pi\, n)}\r)\r]^{-3/2}\r)\;
{\rm sinh}^{2}\,(\sqrt{M}\, \pi\, n)\r]\nn\\
& &\qquad\qquad\qquad\qquad\qquad\qquad\qquad\qquad\quad \times\;
{\rm sinh}^{-3}\, (\sqrt{M}\, \pi\, n)\;\;
{\rm diag.}~(1,1,-2)\nn\\
& &\qquad\qquad\qquad\qquad\qquad\quad\;\, +\l(\frac{3\;
(r^{2}-M\,\ell^{2})}{r^{2}\; {\rm sinh}^{3}\,(\sqrt{M}\, \pi\,
n)}\r)\; \l[1+\l(\frac{M\, \ell^{2}}{r^{2}\; {\rm sinh}^{2}\,
(\sqrt{M}\, \pi\, n)}\r)\r]^{5/2}\nn\\
& &\qquad\qquad\qquad\qquad\qquad\qquad\qquad\qquad\quad \times\;
{\rm diag.}~(1,0,-1). \eea In order to arrive at this finite
result, we have regularized the stress-energy tensor by simply
dropping the $n=0$ term which corresponds to the stress-energy
tensor in AdS$_{3}$. It is well-known that the stress-energy
tensor associated with a massless and conformally coupled scalar
field vanishes in AdS$_{3}$ because of the absence of the trace
anomaly in odd dimensions~\cite{birrell-1982}.

\begin{figure}[!htb]
\begin{center}
\hskip 40pt \resizebox{185pt}{140pt}{\includegraphics{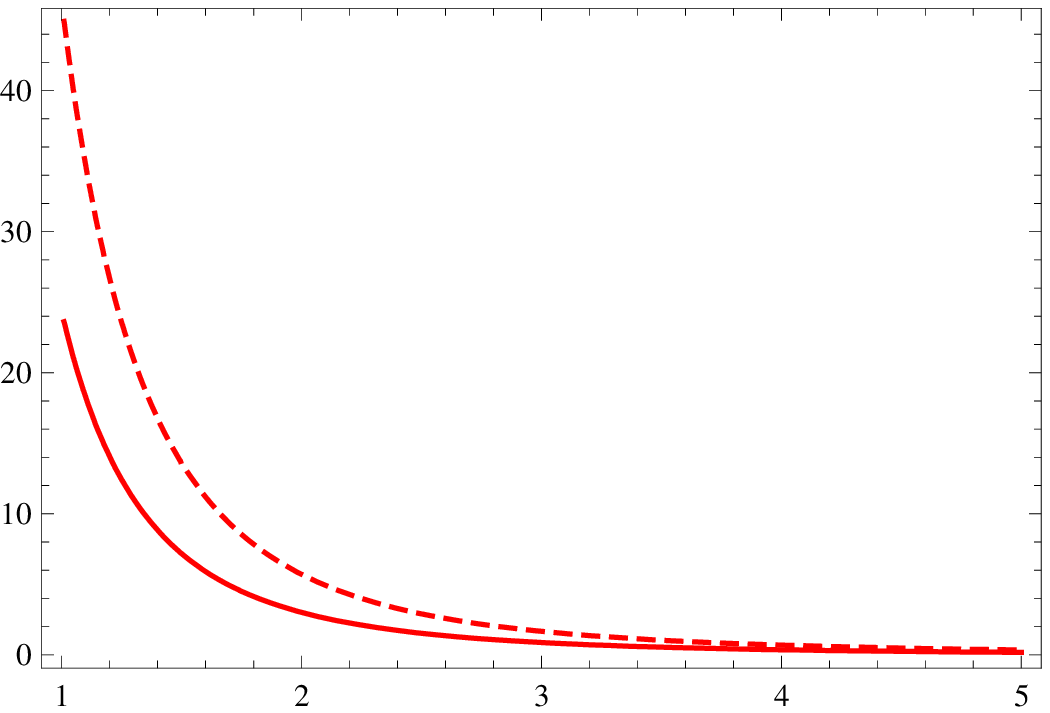}}
\hskip 10pt \resizebox{185pt}{140pt}{\includegraphics{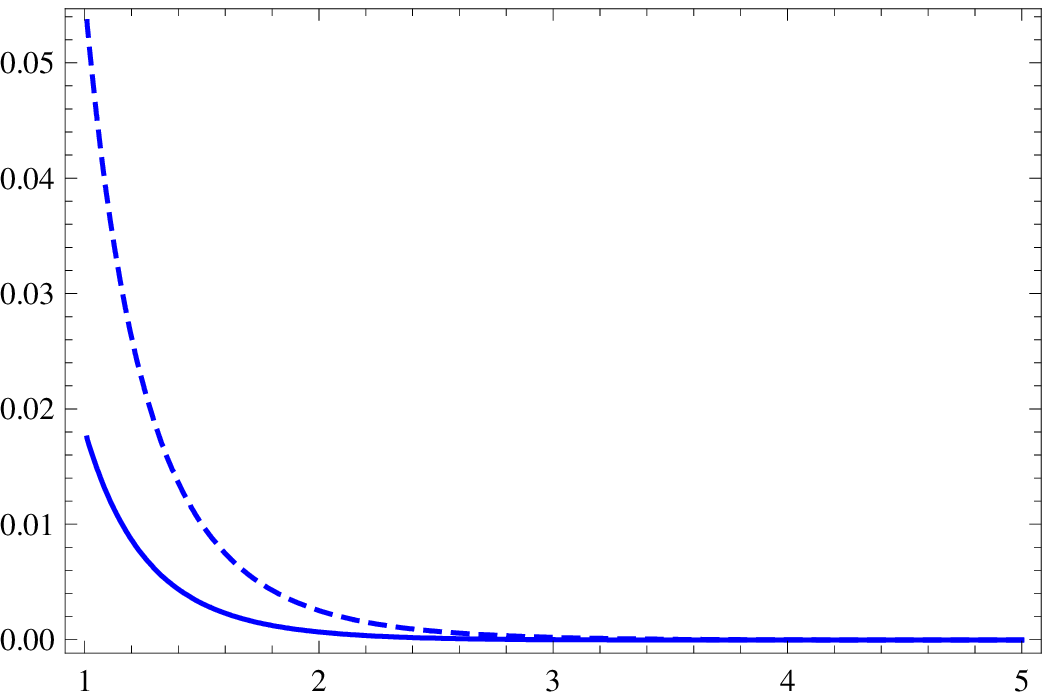}}
\vskip -105 pt \hskip -385 pt $\ell^3\, \langle {\hat
T}^{t}_{t}\rangle$ \vskip 100 pt \hskip 40pt
\resizebox{185pt}{140pt}{\includegraphics{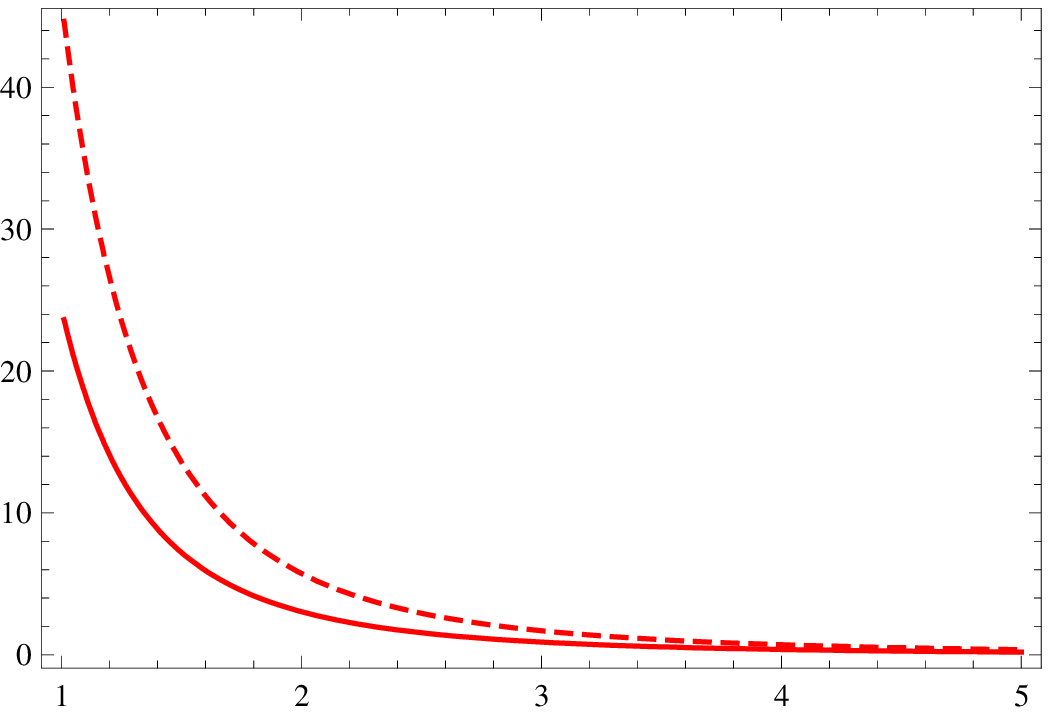}} \hskip 10pt
\resizebox{185pt}{140pt}{\includegraphics{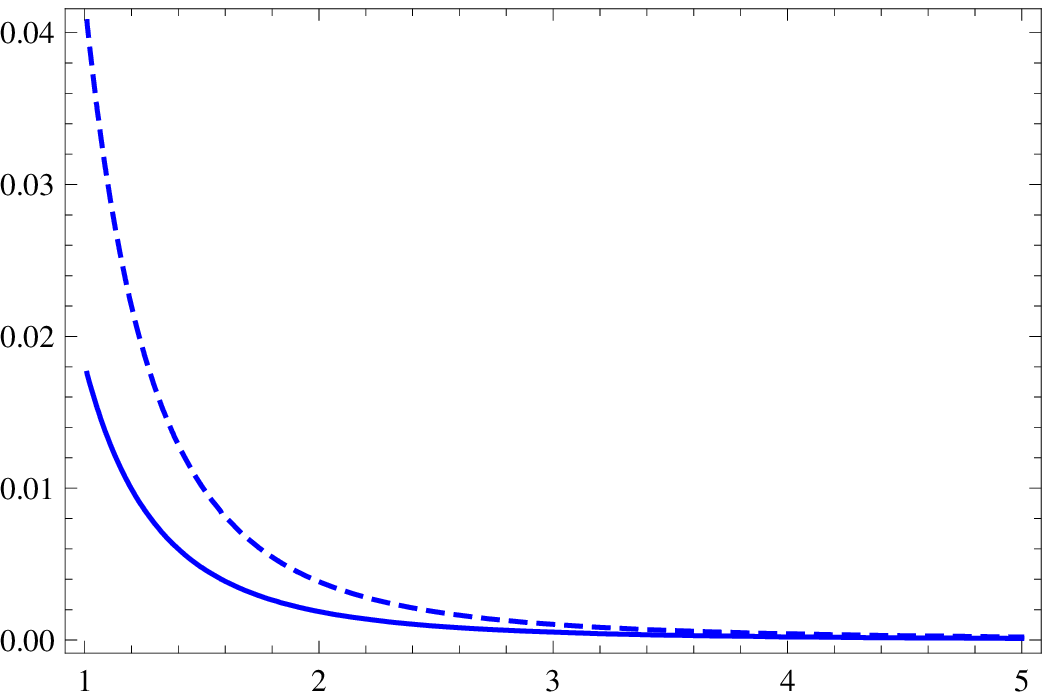}} \vskip -105
pt \hskip -385 pt $\ell^3\, \langle {\hat T}^{r}_{r}\rangle$
\vskip 100 pt \hskip 40pt
\resizebox{185pt}{140pt}{\includegraphics{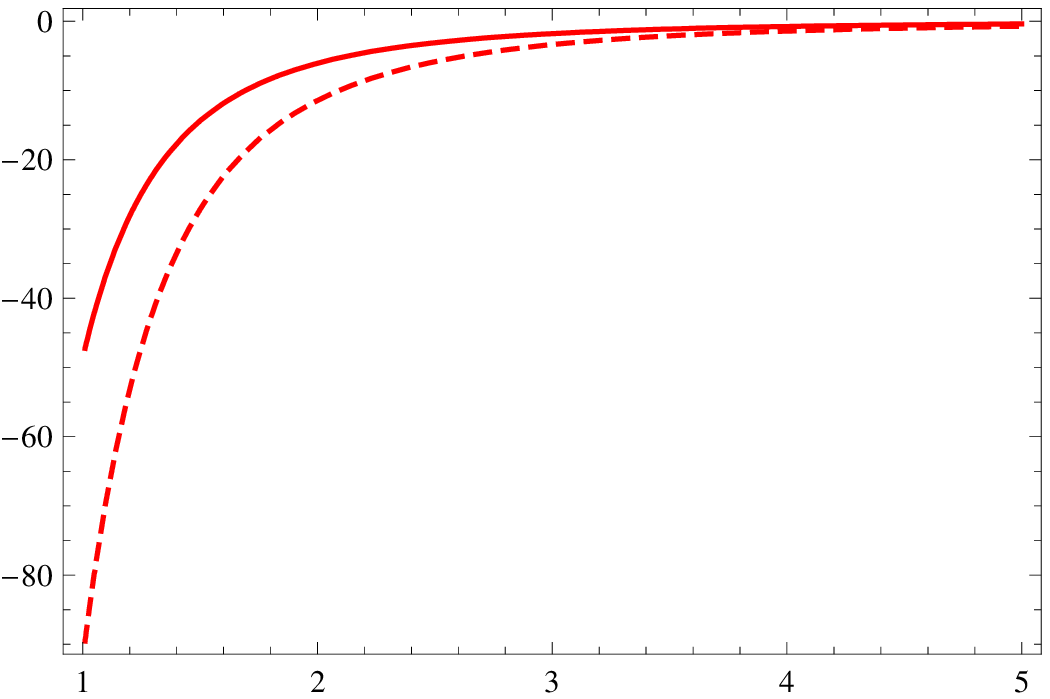}} \hskip 10pt
\resizebox{185pt}{140pt}{\includegraphics{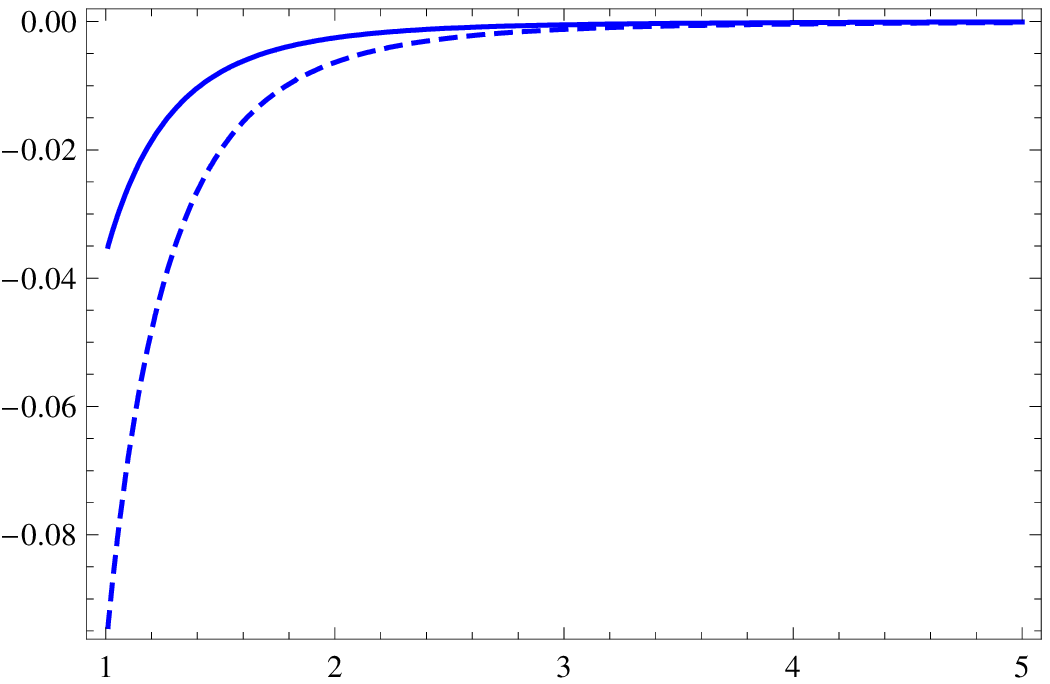}} \vskip -105
pt \hskip -385 pt $\ell^3\, \langle {\hat T}^{\phi}_{\phi}\rangle$
\vskip 90 pt \hskip 45 pt $(r/r_{_{+}})$ \hskip 165 pt
$(r/r_{_{+}})$ \vskip 2 pt \caption{The diagonal components of the
standard (i.e. the unmodified) stress-energy tensor around the BTZ
black hole, viz. $\langle {\hat T}^{t}_{t}\rangle$ (on top),
$\langle {\hat T}^{r}_{r}\rangle$ (in the middle) and  $\langle
{\hat T}^{\phi}_{\phi}\rangle$ (at the bottom), have been plotted
as a function of $(r/r_{_{+}})$. We have set $\ell=10^{3}$ in all
the plots. The red lines (in the left column) and the blue ones
(in the right column) correspond to $M=10^{-3}$ and $M=10^{-1}$,
respectively. In all these plots, the solid lines correspond to
the non-rotating case, while the dashed lines correspond to a
rotating black hole with $(J/M\,\ell)=0.95$. We have chosen such
an extreme value for $J$ to clearly illustrate the effects due to
rotation. It is evident from the above figures that, all the
diagonal components of the stress-energy tensor either increase or
decrease monotically with the distance from the horizon. Also,
qualitatively, rotation does not introduce any new features in
these components, but only changes their magnitude. Moreover, the
larger the~$M$, the smaller is the difference in the magnitude of
these components between the non-rotating and rotating cases.
Furthermore, this difference is the largest at the horizon, and it
decreases with the distance from the horizon of the black hole.}
\label{fig:set1}
\end{center}
\end{figure}

\begin{figure}[!htb]
\begin{center}
\resizebox{280pt}{210pt}{\includegraphics{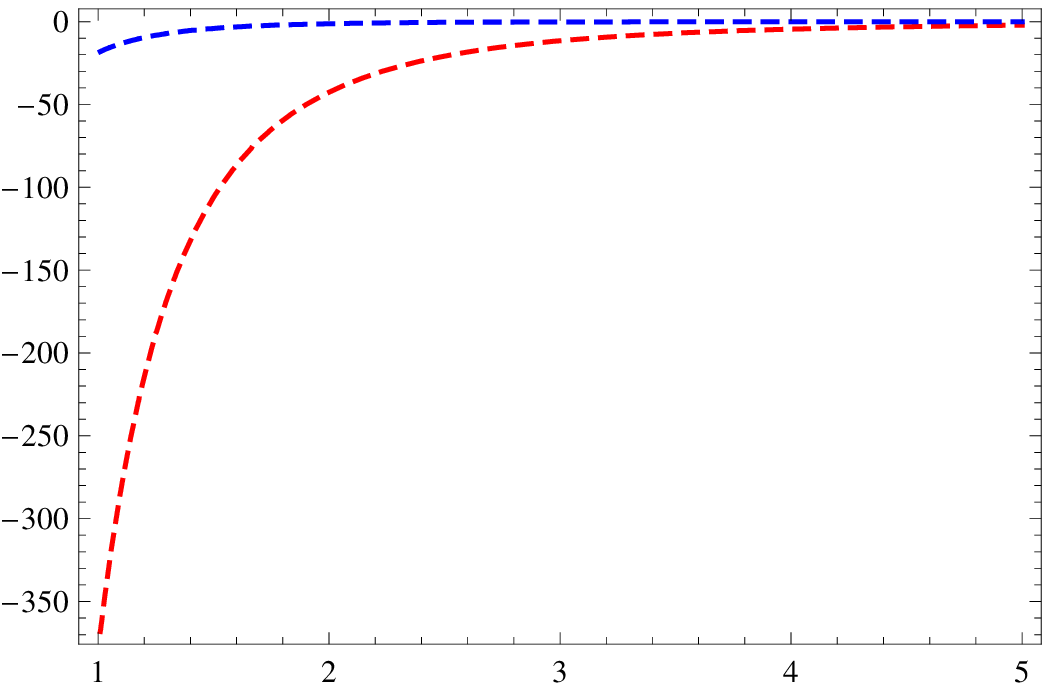}} \vskip -140 pt
\hskip -325 pt $\ell^3\, \langle {\hat T}^{t}_{\phi}\rangle$
\vskip 125 pt \hskip 15 pt $(r/r_{_{+}})$ \vskip 2 pt \caption{The
only non-vanishing, non-diagonal, component of the standard
stress-energy tensor, viz. $\langle {\hat T}^{t}_{\phi} \rangle$,
has been plotted as a function of $(r/r_{_{+}})$. Note that this
component is zero in the non-rotating case. As in the earlier
plots, we have set $\ell=10^{3}$ and chosen $(J/M\,\ell)=0.95$.
The red and blue dashed lines again correspond to $M=10^{-3}$ and
$M=10^{-1}$, respectively. Evidently, the comments that we had
made in the previous figure about the behavior of the diagonal
components of the stress-energy tensor apply to this component as
well.} \label{fig:set2}
\end{center}
\end{figure}

On following the same steps as above, one can, in principle,
obtain an analytic expression for the stress-energy tensor around
the rotating BTZ black hole (i.e. when $J \ne 0$). The
regularization procedure remains the same as in the non-rotating
case---one simply drops the $n=0$ term. When the black hole is
rotating, one finds that, in addition to the diagonal components,
viz. $\langle {\hat T}^{t}_{t} \rangle$, $\langle {\hat
T}^{r}_{r}\rangle$ and  $\langle {\hat T}^{\phi}_{\phi}\rangle$,
the $\langle {\hat T}^{t}_{\phi} \rangle$ component turns out to
be non-zero as well~\cite{steif-1994}. While the procedure for
calculating the stress-energy tensor is rather straightforward,
the resulting expressions turn out to be fairly long\ to be
displayed\footnote{We have evaluated the stress-energy tensor
using {\sl Mathematica}\/~\cite{mathematica}. While we are able to
obtain an unsimplified, analytic expression for the stress-energy
tensor in the rotating case, the expression proves to be rather
long and quite cumbersome. We feel that displaying such an
expression may not particularly aid in visualizing the effects of
rotation on the stress-energy tensor. We shall instead plot all
the components of the stress-energy tensor.}. Therefore, in order
to illustrate the behavior of the stress-energy tensor, in
Figures~\ref{fig:set1} and \ref{fig:set2}, we have plotted all its
components for a couple of different values of the black hole
parameters $M$ and $J$.
Also, to unambiguously demonstrate the effects of rotation, in
addition to the non-rotating case, we have plotted the components
of the stress-energy tensor for a rotating black hole with an
extremely large angular momentum $J$ [we have chosen $(J/M\,\ell)
=0.95$] in the two figures. These figures clearly indicate that,
rotation only changes the magnitude of the stress-energy tensor
and does not alter its qualitative behavior. It either increases
or decreases monotonically with the distance from the horizon, as
in the non-rotating case. Moreover, the larger the $M$, the
smaller is the difference in the stress-energy tensor between the
non-rotating and rotating cases. Furthermore, it is evident from
the figures that, this difference is the maximum at the horizon
and it decreases with the distance from the horizon.

%%%%%%%%%%%%%%%%%%%%%%%%%%%%%%%%%%%%%%%%%%%%%%%%%%%%%%%%%%%%%%%%%%%%%%%%%%%%%%%

\subsection{Planck scale modifications}

\begin{figure}[!htb]
\begin{center}
\hskip 40pt
\resizebox{185pt}{140pt}{\includegraphics{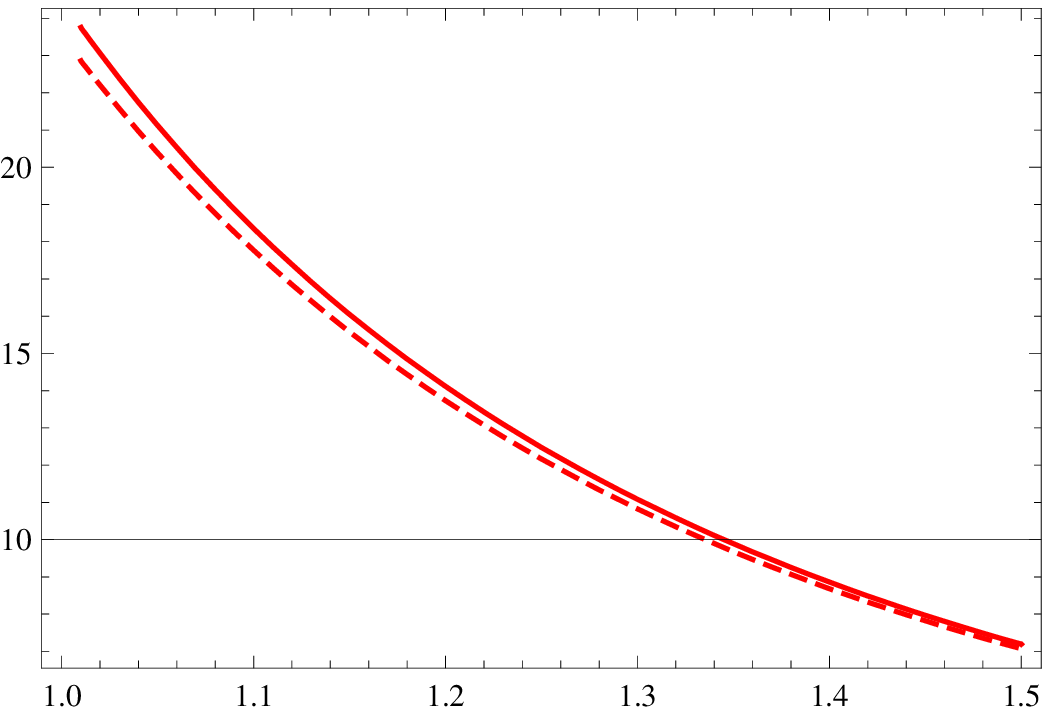}} \hskip
10pt \resizebox{185pt}{140pt}{\includegraphics{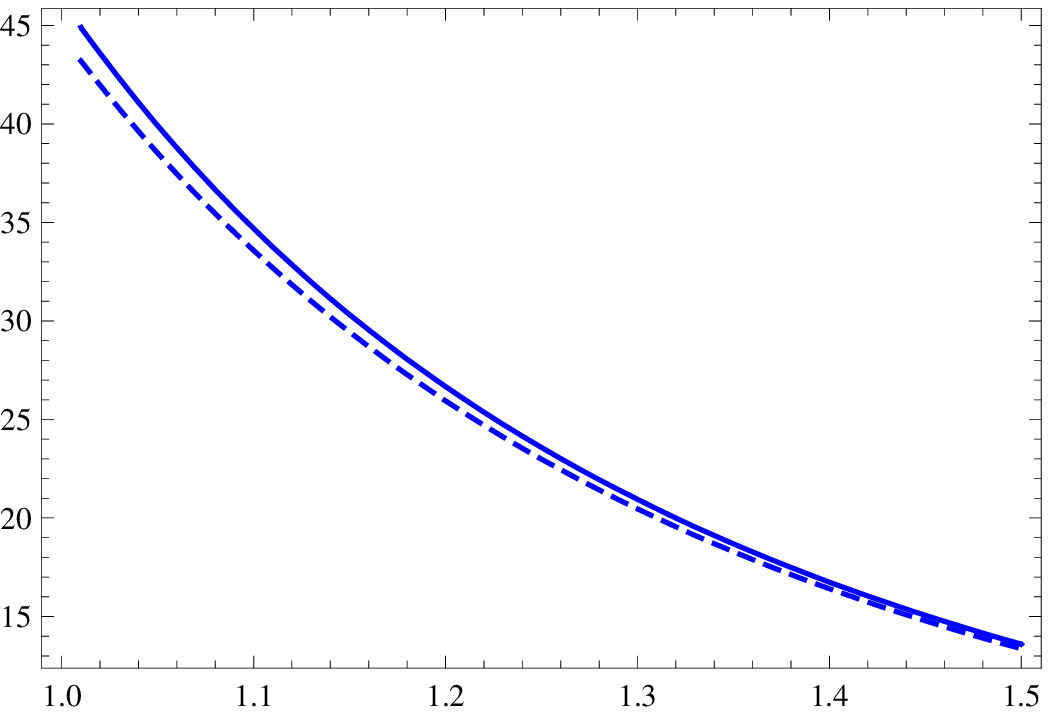}}
\vskip -105 pt \hskip -390 pt $\ell^3\, \langle {\hat T}^{t}_{t}
\rangle_{_{\rm M}}$ \vskip 100 pt \hskip 40pt
\resizebox{185pt}{140pt}{\includegraphics{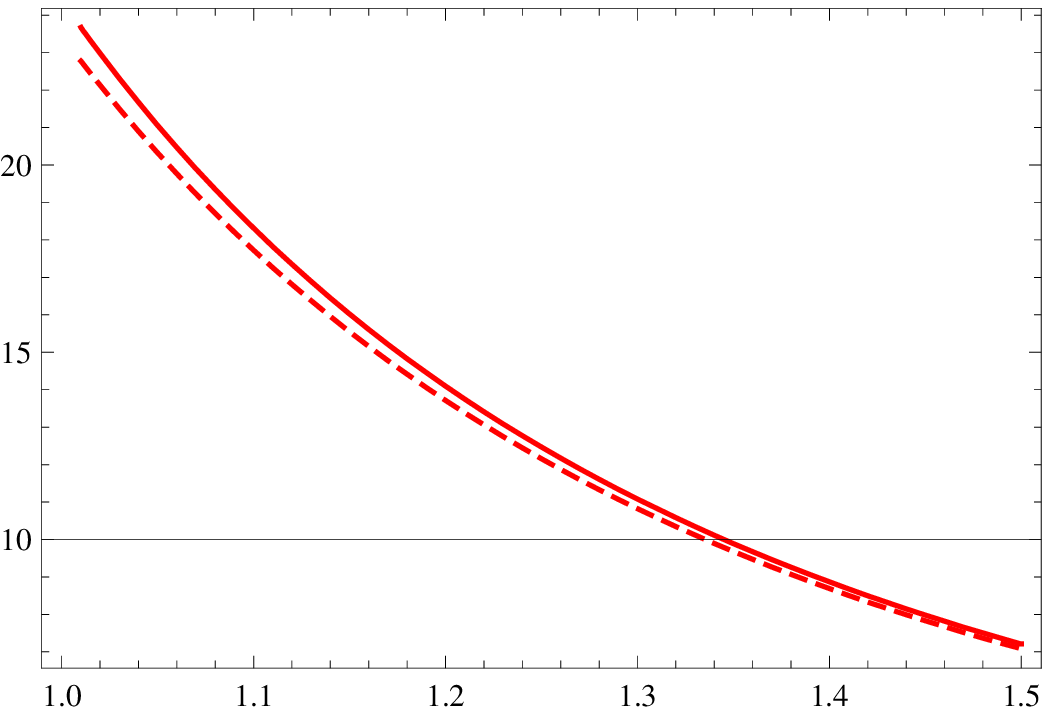}} \hskip
10pt \resizebox{185pt}{140pt}{\includegraphics{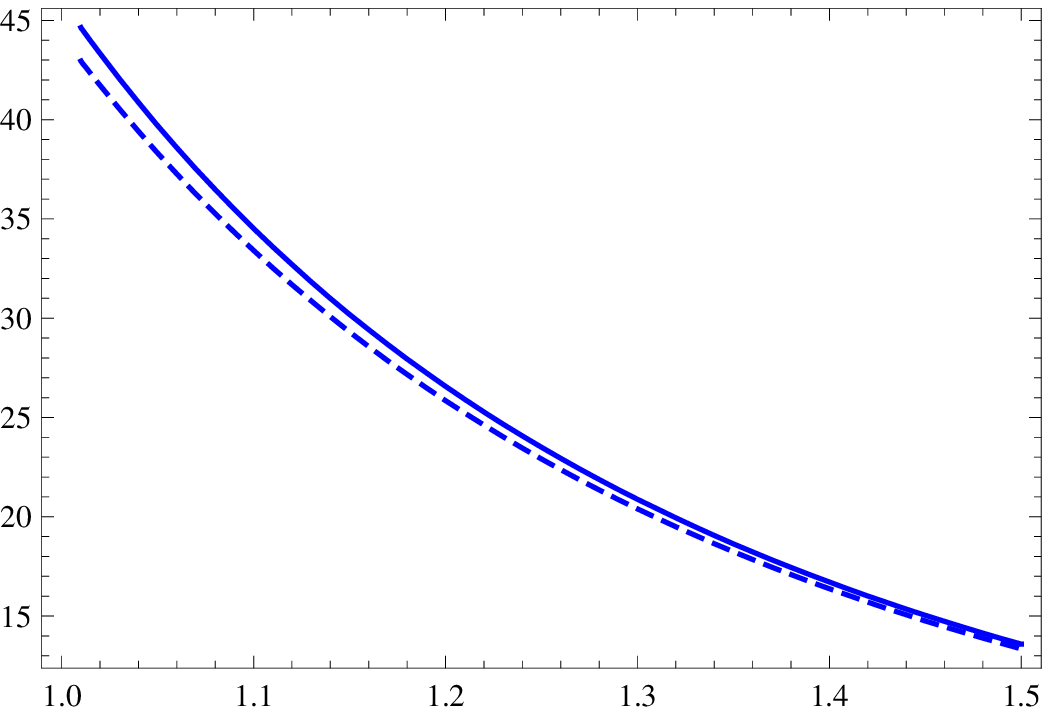}}
\vskip -105 pt \hskip -390 pt $\ell^3\, \langle {\hat T}^{r}_{r}
\rangle_{_{\rm M}}$ \vskip 100 pt \hskip 40pt
\resizebox{185pt}{140pt}{\includegraphics{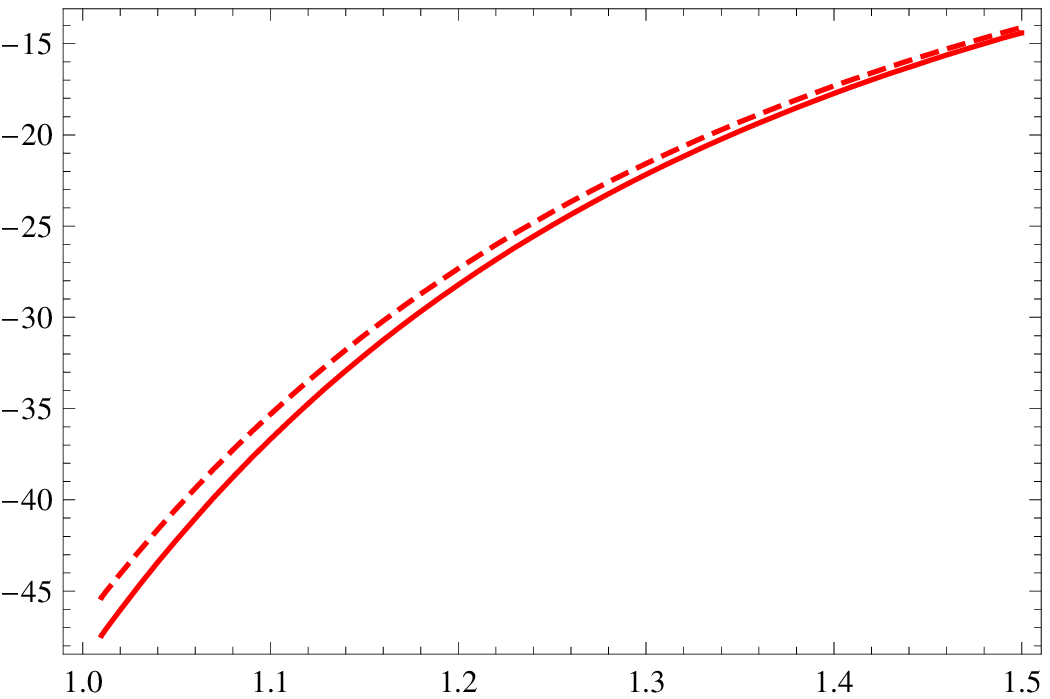}} \hskip
10pt \resizebox{185pt}{140pt}{\includegraphics{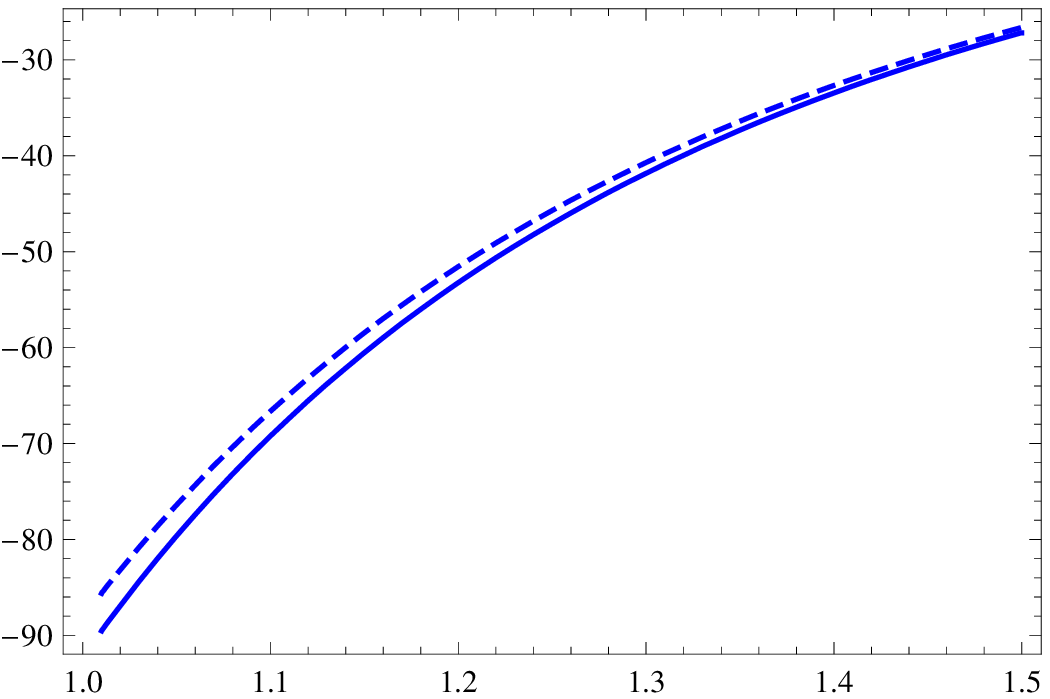}}
\vskip -105 pt \hskip -390 pt $\ell^3\, \langle {\hat
T}^{\phi}_{\phi} \rangle_{_{\rm M}}$ \vskip 90 pt \hskip 45 pt
$(r/r_{_{+}})$ \hskip 165 pt $(r/r_{_{+}})$ \vskip 2 pt
\caption{The diagonal components of the {\it modified}\/
stress-energy tensor around the BTZ black hole, viz. $\langle
{\hat T}^{t}_{t}\rangle_{_{\rm M}}$ (at the top), $\langle {\hat
T}^{r}_{r}\rangle_{_{\rm M}}$ (in the middle) and $\langle {\hat
T}^{\phi}_{\phi}\rangle_{_{\rm M}}$ (at the bottom), have been
plotted as a function of $(r/r_{_{+}})$. We have also plotted the
corresponding standard components for comparison. We have set
$\ell=10^{3}$ and $M=10^{-3}$ in all the plots. The red lines (in
the left column) and the blue ones (in the right column)
correspond to the non-rotating black hole and a rotating one with
$(J/M\,\ell)=0.95$, respectively. In all these plots, the solid
lines correspond to the standard case wherein $\lp=0$, while the
dashed lines correspond to the modified case with $(\lp
/r_{_{+}})=0.9$. We have chosen such an extremely large value of
$\lp$ and have also plotted over a smaller range in $(r/r_{_{+}})$
in order to distinctly demonstrate the Planck scale effects. The
above figures clearly indicate that the Planck scale effects do
not alter the qualitative behavior of the diagonal components of
the stress-energy tensor. Note that the Planck scale modifications
are the largest at the horizon and these modifications decrease
monotonically with the distance from the horizon.}
\label{fig:set3}
\end{center}
\end{figure}

\begin{figure}[!htb]
\begin{center}
\resizebox{280pt}{210pt}{\includegraphics{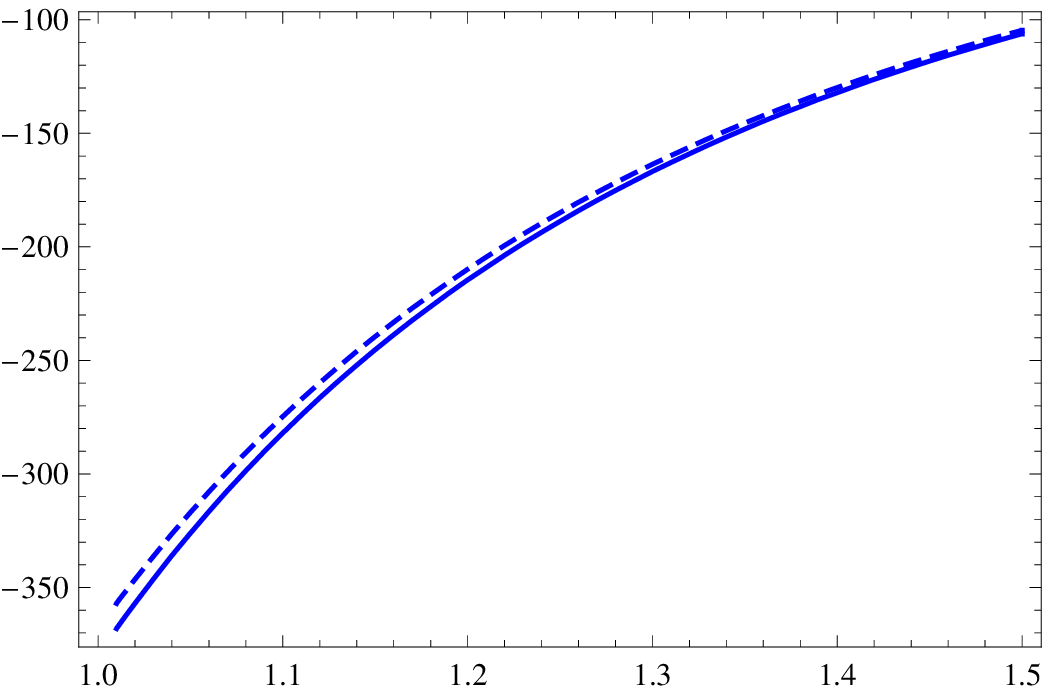}} \vskip -140
pt \hskip -325 pt $\ell^3\, \langle {\hat T}^{t}_{\phi}
\rangle_{_{\rm M}}$ \vskip 125 pt \hskip 15 pt $(r/r_{_{+}})$
\vskip 2 pt \caption{The only non-zero and non-diagonal component
of the {\it modified}\/ stress-energy tensor, viz. $\langle {\hat
T}^{t}_{\phi}\rangle_{_{\rm M}}$, has been plotted as a function
of $(r/r_{_{+}})$. As in the standard case, we find that this
component vanishes in the modified case when the black hole is not
rotating. We have set $\ell=10^{3}$ and chosen $(J/M\,\ell)=0.95$
in these plots. For comparison, we have also plotted the
corresponding component in the unmodified case. The blue solid
line corresponds to the standard case (i.e. when $\lp=0$), while
the blue dashed line corresponds to the modified case with
$(\lp/r_{_{+}})=0.9$. It is evident that the comments that we had
made in the previous figure about the behavior of the diagonal
components of the modified stress-energy tensor apply to this
component as well.}\label{fig:set4}
\end{center}
\end{figure}

The Planck scale modifications to the stress-energy tensor around
the rotating BTZ black hole can now be arrived at upon using the
modified Green's function~(\ref{eq:mGfnBTZ}) in the
expression~(\ref{eq:set}). The structure of the modified
stress-energy tensor turns out to be the same as in the standard
case. When the black hole is not rotating, the only non-zero
components of the modified stress-energy tensor are the diagonal
components, which we shall refer to as $\langle {\hat
T}^{t}_{t}\rangle_{_{\rm M}}$, $\langle {\hat
T}^{r}_{r}\rangle_{_{\rm M}}$ and $\langle {\hat
T}^{\phi}_{\phi}\rangle_{_{\rm M}}$. Around a rotating black hole,
as in the unmodified case, the only additional non-vanishing
component turns out to be  $\langle {\hat
T}^{t}_{\phi}\rangle_{_{\rm M}}$. However, as in case of the
standard stress-energy tensor around a rotating black hole, the
resulting expressions for the modified stress tensor prove to be
rather long. We believe that displaying these long and unwieldy
expressions may not be necessarily helpful in understanding the
Planck scale effects. Therefore, we have again plotted the various
components of the modified stress-energy tensor (along with the
unmodified ones) in Figures~\ref{fig:set3} and~\ref{fig:set4}.

In order to distinctly show the Planck scale modifications, in
these two figures, we have plotted the non-vanishing components of
the stress-energy tensor for an extremely large (and unrealistic)
value of $\lp$ [we have set $(\lp/r_{_{+}})=0.9$]. For convenience
in comparison, we have also plotted the corresponding unmodified
components in these figures.
It is evident from the figures that the Planck scale effects (as
taken into account through the duality principle) do not modify
the stress-energy tensor to any appreciable extent. Nor do they
alter its qualitative behavior. As in the standard case, the
modified stress-energy tensor either increases or decreases
monotonically with distance from the horizon. Moreover, the Planck
scale modifications prove to be significant only very close to the
horizon. We find that, for an $\lp$ as large as $(0.9\;
r_{_{+}})$, the change in the stress-energy tensor is of the order
of $3$--$5\%$ at the horizon. For any smaller value of $\lp$, the
modified case turns out to be completely indistinguishable from
the unmodified one. Therefore, we can conclude that, for any
realistic $\lp$, the modifications are completely negligible. This
result corroborates similar conclusions that have been arrived at
earlier in the literature (see, for example,
Refs.~\cite{jacobson-1993-99,brout-1995-99,hambli-1996,agullo-2007}).

%%%%%%%%%%%%%%%%%%%%%%%%%%%%%%%%%%%%%%%%%%%%%%%%%%%%%%%%%%%%%%%%%%%%%%%%%%%%%%%

\section{Discussion}\label{sec:summary}

In this work, using the T-duality symmetry of the string
fluctuations, we have evaluated the modified two-point function
and the resulting stress-energy tensor around the rotating BTZ
black hole. We should emphasize here that we have not made any
approximation whatsoever in our calculations and the results we
have obtained around the BTZ black hole are exact. This is
important, since, as we had mentioned in the introductory section,
much of the earlier analyses had either worked with the moving
mirror model of Hawking radiation in $(1+1)$-dimensions or had
just considered the spherically symmetric mode in
$(3+1)$-dimensions, which, effectively, simplifies to the lower
dimensional model. Moreover, we are not aware of any earlier
analysis in the literature wherein the Planck scale effects have
been studied around a rotating black hole.

Interestingly, we find that the modified Green's function remains
finite in the coincident limit. Actually, such a result could have
been expected based on the Schwinger-DeWitt expansion of the
kernel $K({\tilde x}, {\tilde x'}; s)$~\cite{dewitt-1975}.
According to the expansion, for small separations, the kernel in
an arbitrary spacetime has the same form as in the Minkowski
vacuum. Therefore, if the modified Green's function is
ultra-violet regulated in flat spacetime, then it can be expected
to remain finite in the coincident limit in any curved spacetime
as well. Further, we find that the Planck scale modifications to
the stress-energy tensor are negligibly small, in agreement with
similar conclusions that have been arrived at earlier in the
literature~\cite{jacobson-1993-99,brout-1995-99,hambli-1996,agullo-2007}.

Ideally, rather than evaluate the Planck scale modifications to
the stress-energy tensor, one would like to evaluate the
corrections to Hawking radiation itself. This in turn requires
that one considers a dynamical situation, and evaluates either the
effective Lagrangian or the in-out Bogoliubov coefficient, while
taking into account the Planck scale modifications. We are
currently investigating these issues.

%%%%%%%%%%%%%%%%%%%%%%%%%%%%%%%%%%%%%%%%%%%%%%%%%%%%%%%%%%%%%%%%%%%%%%%%%%%%%%%

%%%%%%%%%%%%%%%%%%%%%%%%%%%%%%%%%%%%%%%%%%%%%%%%%%%%%%%%%%%%%%%%%%%%%%%%%%%%%%%

\section*{Acknowledgments}

The authors would like to thank T.~Padmanabhan for discussions.
DAK and LS wish to thank the Harish-Chandra Research Institute,
Allahabad, India, and the Inter University Centre for Astronomy
and Astrophysics, Pune, India, respectively, for hospitality,
where part of this work was carried out. DAK is supported by the
Senior Research Fellowship of the Council for Scientific and
Industrial Research, India. SS is supported by the Marie Curie
Incoming International Grant IIF-2006-039205.

%%%%%%%%%%%%%%%%%%%%%%%%%%%%%%%%%%%%%%%%%%%%%%%%%%%%%%%%%%%%%%%%%%%%%%%%%%%%%%%
%%%%%%%%%%%%%%%%%%%%%%%%%%%%%%%%%%%%%%%%%%%%%%%%%%%%%%%%%%%%%%%%%%%%%%%%%%%%%%%

%%%%%%%%%%%%%%%%%%%%%%%%%%%%%%%%%%%%%%%%%%%%%%%%%%%%%%%%%%%%%%%%%%%%%%%%%%%%%%%
%%%%%%%%%%%%%%%%%%%%%%%%%%%%%%%%%%%%%%%%%%%%%%%%%%%%%%%%%%%%%%%%%%%%%%%%%%
\end{document}